Depressive patients are more impulsive and inconsistent in intertemporal choice behavior for monetary gain and loss than healthy subjects- an analysis based on Tsallis' statistics.

Taiki Takahashi[1], Hidemi Oono[2], Takeshi Inoue[3], Shuken Boku[3], Yuki Kako[3], Yuji Kitaichi[3], Ichiro Kusumi[3], Takuya Masui[3], Shin Nakagawa[3], Katsuji Suzuki[3], Teruaki Tanaka[3], Tsukasa Koyama[3], and Mark H. B. Radford[4]

[1]Direct all correspondence to Taiki Takahashi, Unit of Cognitive and Behavioral Sciences Department of Life Sciences, School of Arts and Sciences, The University of Tokyo, Komaba, Meguro-ku, Tokyo, 153-8902, Japan (taikitakahashi@gmail.com).

[2]Department of Behavioral Science, Faculty of Letters, Hokkaido University, N.10, W.7, Kita-ku, Sapporo, 060-0810, Japan

[3]Department of Psychiatry, Graduate School of Medicine, Hokkaido University, N.15, W.7, Kita-ku, Sapporo, 060-8638

[4]Symbiosis Group Limited, P.O. Box 1192, Milton, 4064 Australia; and Department of Behavioral Science, Hokkaido University, N.10, W.7, Kita-ku, Sapporo, 060-0810, Japan

Acknowledgements: The research reported in this paper was supported by a grant from the Grant- in-Aid for Scientific Research ("21st century center of excellence" grant) from the Ministry of Education, Culture, Sports, Science and Technology of Japan.

**Abstract**

Depression has been associated with impaired neural processing of reward and punishment. However, to date, little is known regarding the relationship between depression and intertemporal choice for gain and loss. We compared impulsivity and inconsistency in intertemporal choice for monetary gain and loss (quantified with parameters in the q-exponential discount function based on Tsallis' statistics) between depressive patients and healthy control subjects. This examination is potentially important for advances in neuroeconomics of intertemporal choice, because depression is associated with reduced serotonergic activities in the brain. We observed that depressive patients were more impulsive and time-inconsistent in intertemporal choice action for gain and loss, in comparison to healthy controls. The usefulness of the q-exponential discount function for assessing the impaired decision-making by depressive patients was demonstrated. Furthermore, biophysical mechanisms underlying the altered intertemporal choice by depressive patients are discussed in relation to impaired serotonergic neural systems.

# 1. Introduction
## 1.1 Impulsivity and depression

Depressive patients have neuropsychological impairments including deficits in sensitivity to reward and punishment [1] and decision-making [2]. It is also known that altered serotonergic neural systems are associated with depression and related decision-making impairments [3]. Because depressive disorders are strongly associated with impulsive and risky behavior such as suicide attempts [4,5], it is of importance to establish a rigorous behavioral framework for assessing the degrees of impairments in decision-making by depressive subjects. In this study, we examined two distinct neuropsychological tendencies; i.e., impulsivity and inconsistency in intertemporal choice (delay discounting) by clinically diagnosed depressive patients. Briefly, impulsivity in intertemporal choice refers to the degree of preference for (or aversion to) smaller sooner rewards (punishments) over later larger ones; while inconsistency in intertemporal choice refers to time-dependency of the impulsivity in intertemporal choice (we will illustrate these two tendencies in the next section). Because previous studies indicate that serotonergic functioning in the brain (known to be reduced in depressed patients) may be related to the evaluation of future rewards [6], these investigations are of potential importance for neuroeconomic understandings of depression and biophysical mechanisms of serotonergic systems underlying impaired decision-making by mood disorder patients. However, to date, little is known about the relationships between depression, impulsivity and inconsistency in intertemporal choice, partly due to a difficulty in operationalizing impulsivity and inconsistency in intertemporal choice in a distinct manner.

Recent studies in econophysics and neuroeconomics have demonstrated the usefulness of the q-exponential discount function [7-9] to parametrize both impulsivity and inconsistency in intertemporal choice. We therefore compared both impulsivity and inconsistency in intertemporal choice for monetary gains and losses between depressive patients and healthy control subjects, by utilizing the q-exponential discount function. It is further to be noted that this study is the first to compare both impulsivity and time-inconsistency in intertemporal choice between control and neuropsychiatric patient groups by utilizing the q-exponential discount function based on Tsallis' statistics.

## 1.2 Impulsivity and inconsistency in intertemporal choice

Impulsivity in intertemporal choice (delay discounting) for gain refers to preference for smaller but more immediate rewards over larger but more delayed ones. Consider the following example:

(A) Choose between (A.1) One cup of coffee now.
                         (A.2) Two cups of coffee tomorrow.
(B) Choose between (B.1) One cup of coffee in one year.
                         (B.2) Two cups of coffee in [one year plus one day].

Most people may prefer a sooner smaller reward in (A) (i.e., (A.1), impulsive choice); while prefer a later larger reward in (B) (i.e., (B.2), patient choice). These examples demonstrate that most people are patient in making a plan about intertemporal choice in the distant future, but impulsive in the near future, resulting in "preference reversal" as time passes [10-12]. It is to be noted that consistent decision-makers should choose either [(A1) and (B1)] or [(A2) and (B2)] because time-intervals between sooner smaller rewards and later larger ones are the same (i.e., one day) in both (A) and (B). In summary, impulsivity in delay discounting of gain corresponds to the degree to which the subject discount the delayed reward; while inconsistency in delay discounting corresponds to the dependency of the intensity of aversion to waiting for one day to obtain an additional cup of coffee, on time-points (now or 1 year later) in the examples above. It is also to be noted that example A is about the actual intertemporal choice *action*; on the other hand, example B is about the intertemporal choice *plan*. Note that people cannot actually take future actions now, and therefore choosing (B2) is a future *plan* (not an actual action); while choosing (A1) is an actual *action*. Neuroeconomic studies have reported that addiction to drugs of abuse is associated with impulsive intertemporal choice for gain (*e.g.* choosing immediate rewards from drug intake at the cost of later larger rewards such as healthy body in later life) [13-16]; while little is known about the relationships between neuropsychiatric illnesses including addiction and inconsistency in intertemporal choice.

      In intertemporal choice for loss, discounting of delayed loss corresponds to a decrease in aversion to loss when the loss is delayed. In other words, subjects who avoid paying small costs immediately and choose to pay larger later costs are strong delay discounters of loss. Therefore, strong delay discounting of loss (impulsivity in delay discounting of loss) represents the marked tendency of procrastination about paying a cost.

**1.3 q-exponential discount function based on Tsallis' statistics**
      Recent econophysical and neuroeconomic studies [7-9] proposed and examined the following q-exponential discount function based on Tsallis' statistics:

$$V(D)=A/ exp_q(k_qD)=A/[1+(1-q)k_qD]^{1/(1-q)} \quad \text{(Equation 1)}$$

where $exp_q$ () is the q-exponential function, D is a delay until receipt of a reward, $A$ is the value of a reward at $D=0$, and $k_q$ is a parameter of impulsivity at delay $D=0$. Note that when $q=0$, equation 1 is the same as a simple hyperbolic discount function [10-12]:

$V(D)=A/(1+k_hD),$ (Equation 2),

where $k_h=k_0$, while $q \to 1$, is the same as an exponential discount function proposed in classical economics [17]:

$V(D)=Aexp(-k_eD),$ (Equation 3)

where $k_e=k_1$.

In any continuous time-discounting functions, a discount rate is defined as $-(dV(D)/dD)/V(D)$, independently of functional types of discount models, and larger discount rates corresponds to more impulsive intertemporal choice. In the q-exponential discount function, the discount rate (q-exponential discount rate) is:

$-V'(D)/V(D)=k_q/(1+k_q(1-q)D).$ (Equation 4)

We can see that when $q=1$, the discount rate is independent of delay $D$, corresponding to exponential discounting (consistent intertemporal choice); while for $q<1$, the discount rate is a decreasing function of delay $D$, resulting in preference reversal. This can also be demonstrated by calculating the time-derivative of the q-exponential discount rate:

$(d/dD) [-V'(D)/V(D)]= -k_q^2(1-q)/(k_q(1-q)D+1)^2$ (Equation 5)

which is negative and positive for $q<1$ and $q>1$, respectively. Also, impulsivity at delay D=0 is equal to $k_q$ irrespective of $q$. We have previously shown that the q-exponential discount function is capable of continuously quantifying human subjects' inconsistency in intertemporal choice [8,9]. Namely, human agents with smaller q values are more inconsitent in intertemporal choice. If $q$ is less than 0, the intertemporal choice behavior is more inconsistent than hyperbolic discounting.

**1.4 Objectives of the present study**

The aim of this study was to examine impulsivity ($k_q$ in Equation 1) and inconsistency (Equation 2) in delay discounting of gain and loss among depressed patients, in comparison to healthy people. Based on previous research that has suggested higher impulsivity among depressed patients, we propose the hypothesis that the depressed patients are more impulsive in intertemporal choice behavior than healthy normal subjects. It is important to note that our present study is the first to compare inconsistency in intertemporal choice between healthy controls and neuropsychiatric patients. Notably, previous findings regarding the relationship between serotonin and discounting have been mixed. A rodent animal model reported that a reduction in serotonergic activities were associated with exaggerated impulsivity in intertemporal choice over several seconds [18]; while a psychopharmacological study with human subjects did not observe a significant effect of a reduction in serotonergic activities on intertemporal choice over a year [19]. How our present study resolves the discrepancy between these studies is also discussed after presentation of our experimental data.

Furthermore, we also aimed to examine the differences in impulsivity and inconsistency between discounting delayed gain and loss. The rationale is that no study to date examined the effect of the sign of delayed outcomes (i.e., gain or loss) on inconsistency in delay discounting, although it has been reported that delayed gain is more rapidly discounted than delayed loss (i.e., the sign effect in intertemporal choice [12]).

## 2. Methods
### 2.1 Participants

Participants were 29 depressive patients diagnosed with DSM-IVTR (major depressive disorder and bipolar disorder, most recent episode: depressed) and 15 healthy control subjects. The depressive patient participants were patients of Hokkaido University Hospital, which serves a large urban catchments area. Participants' ages ranged from 27 to 67, with a mean age of 43.83 years (SD = 2.25), and healthy controls' ages ranged from 31 to 71 with a mean age of 47.6 years (SD = 3.53). All participants including depressive patients and healthy control subjects filled an informed consent form before starting the experiment. The group of depressive patients consisted of both unipolar (i.e., patients with only a major depression phase) and bipolar (i.e., patients with both depressive and manic phases) disorders. It is important to note that bipolar disorder patients in a depressed state at the time of their participation (not in a manic state) were included in the present study, in order to exclude the influences of manic mental states on intertemporal choice. Consequently, there was no significant difference in intertemporal choice behavior between unipolar and bipolar disorder patients. Therefore, we combined both unipolar and bipolar patients (referred to as "depressive patients", hereafter). Moreover, most patients with depression were under medical treatment with several types of antidepressants/mood stabilizers. We did not, however, find any significant effect of the types of antidepressants/mood stabilizers on intertemporal choice. Therefore, we did not divide the patients into sub-groups according to the types of antidepressants they intake. The characteristics of these subjects are described in more detail in Table 1. Subjects with past or current illegal drug use, assessed with Mini-International Neuropsychiatric Interview (MINI, see below), were excluded from the present study, in order to avoid the influences of substance abuse on intertemporal choice.

**2.2 Intertemporal choice task**

As a standard instrument measuring degrees to which participants discount delayed reward and loss, we conducted a face-to-face task developed by neuropsychopharmacologists Bickel and colleagues [20]. It is to be noted that we have also utilized this task in our previous studies [8,9,21].

First, participants were seated individually in a quiet room, and faced the experimenter across a table. After that, participants received the simple instruction that monetary rewards (or losses) in this experiment were hypothetical, but the experimenter wanted them to think as though they were real money. Then the participants were asked to choose between the card describing money delivered immediately (or paid immediately, in the loss condition) and the card describing money delivered after certain delay (or paid after a certain delay, in the loss condition). The left card viewed by participants indicated the amounts of money that could be received immediately (or that had to be paid immediately, in the loss condition), and the right card indicated 100,000 yen that could be received after a certain delay (or that had to be paid after a certain delay, in the loss condition.

For the delay discounting tasks, monetary rewards (or losses) and the delay time were printed on 3×5 index cards. The 27 monetary amounts were 100,000 yen (about $1,000), 99,000 yen, 96,000 yen, 92,000 yen, 85,000 yen, 80,000 yen, 75,000 yen, 70,000 yen, 65,000 yen, 60,000 yen, 55,000 yen, 50,000 yen, 45,000 yen, 40,000 yen, 35,000 yen, 30,000 yen, 25,000 yen, 20,000 yen, 15,000 yen, 10,000 yen, 8,000 yen, 6,000 yen, 4,000 yen, 2,000 yen, 1,000 yen, 500 yen and 100 yen. The seven time delays were 1 week, 2 weeks, 1 month, 6 months, 1 year, 5 years and 25 years.

The experimenter turned the 27 cards sequentially. The card started with 100 yen, up to 100,000 yen in the ascending order condition, or started with 100,000 yen, down to 100 yen, in the descending order condition. For each card, participants chose either the immediate or the delayed reward (or loss). The experimenter wrote down the first delayed reward chosen in the descending gain condition, first immediate reward chosen in the ascending gain condition, first immediate loss in the descending loss condition, and first delayed loss in the ascending loss condition. The average of these results (gain condition and loss condition were calculated separately), were used as the points or subjective equality (hereafter called the *indifference point*, a subjective value of delayed gain/loss) in the following analyses. This procedure was repeated for each of the seven delays (for more detail, see Bickel et al., 1999). The four conditions (ascending gain condition, descending gain condition, ascending loss condition, and descending loss condition) were conducted randomly for each participant. These conditions of the order of the delay discounting tasks did not significantly influence the results. Because our aim was to compare the group difference between healthy controls and depressive patients, we presented estimated parameters of the q-exponential discounting (see the next section) for group median data of the indifference point at each delay. It should also be noted that when the analysis was performed at the individual level, essentially the same conclusions were obtained.

After the determination of indifference points in the delay discounting tasks, we estimated parameters for the q-exponential discount function (i.e., $k_q$ and $q$ in Equation 1, corresponding to impulsivity and inconsistency, respectively) for gain and loss, separately. For estimating the parameters, we conducted nonlinear curve fitting with the Gauss-Newton algorithm implemented in R statistical language (nonlinear modeling package).

**2.3 Questionnaires**

*Mini-International Neuropsychiatric Interview (MINI)*

To examine the participants' drug abuse and alcohol abuse, we used MINI [22]. This scale is also a structured diagnostic scale, administered by well-trained psychiatric doctors when interviewing participants. Although this scale can divided into several parts for assessing specific psychiatric disorders, we used only 2 parts – those for assessing drug abuse and alcohol abuse.

*Beck Depression Inventory-II (BDI-II)*

To assess both healthy and depressive participant's degrees of depressive tendency, we assessed Beck's Depression Inventory (more specifically, BDI-II) [23,24]. This scale is a commonly used self-report scale that measures severity of depression 'over the past week'. The scale consists of 21 items that describe core depressive symptoms, with score being rated on a 4-point scale.

In this study, we used the Japanese version of the BDI-II [25,26].

*Demographic questionnaire*

In addition to the above measures, participants completed questionnaires about sex, age, history of smoking, and suicide attempt histories. These demographic data did not significantly affect intertemporal choice behavior in the present study. Therefore, we simply compared parameters in the q-exponential discount function for gain and loss between healthy controls and depressed patients.

**2.4 Data analysis**

For statistically testing differences in estimated parameters in the q-exponential discount function at the individual level, we utilized t-tests. It is also to be noted that we examined the fitness of the discount models with AICc (Akaike Information Criterion with small sample correction) and observed that the q-exponential discount function best fitted among other models (i.e., exponential and hyperbolic functions) in both healthy controls and depressive patients.

All statistical procedures were conducted with R statistical language (http://www.r-project.org/).　Significant level was set at 5% throughout.

**3. Results**
**3.1 Impulsivity and inconsistency in controls and depressive patients**
Demographic characteristics of depressive patients and healthy controls are presented in Table 1. We observed that BDI-II scores were higher for depressive patients than controls ($p<0.05$), verifying that our present populations were appropriate for the objectives of the present study.

The estimated parameters in the q-exponential discount function for gain and loss at the group level are summarized in Table 2 (Table 2-1 for gain, Table 2-2 for loss). Fitted q-exponential curves for group median indifference points were presented in Fig. 1 (Fig. 1A for gain, Fig.2B for loss). We can see that depressive patients less dramatically discounted delayed outcomes in the distant future, compared to healthy controls, implying that depressive patients make more patient (less impulsive) intertemporal choice *plan*s in the distant future. However, the patient plan does not imply that their actual intertemporal choice *action* is also patient (less impulsive), because it is possible for preference reversal to occur due to inconsistency in intertemporal choice (see section 1.2). In order to investigate differences in both impulsivity at delay D=0, (i.e., impulsivity in intertemporal choice *action*) and inconsistency in delay discounting of gain and loss by controls and depressive patients at the individual level, we conducted t-tests on estimated $k_q$ (impulsivity at delay D=0) and *q* (time-consistency) between controls and depressive patients. Consequently, we found that depressive patients had significantly larger $k_q$ (impulsivity at delay D=0) and smaller *q* (time-consistency) for both gain and loss, in comparison to controls ($p$s<0.05). In other words, depressive patients were more impulsive and inconsistent in delay discounting *action*s for both gain and loss than healthy control subjects (see Table 2, for group data). Moreover, as noted earlier, there was no significant effect of demographic variables other than depressive status (e.g. sex, age, histories of suicide attempts, and the type of antidepressants) on the estimated parameters in the q-exponential discount function.

### 3.2 Gain-loss asymmetries

Next, we plotted the estimated q-exponential discount rate defined in Equation 4 (for gain and loss) of healthy controls and depressive patients (Fig. 2). As can be seen from Fig.2, depressive patients were more impulsive (i.e., a larger discount rate) in the near future, but less impulsive (i.e., a smaller discount rate) in the distant future, in comparison to healthy controls. This indicates that depressive patients may experience more exaggerated "preference reversal" in their intertemporal choice between *plan*s and *action*s. It is important to note that depressive patients' discount rate for future loss at the delay of several decades is almost zero, indicating that depressive patients had hypersensitivity to potential bad outcomes in the extremely distant future (a red curve, Fig 2B).

In order to examine differences in impulsivity and inconsistency in intertemporal choice for gain and loss at the individual level, we conducted t-tests on estimated $k_q$ (impulsivity at delay D=0) and *q* (time-consistency) between gain and loss. We then observed that $k_q$ was significantly smaller for gain; while *q* was larger for gain ($p$<0.05), indicating that intertemporal choice for loss is less impulsive (in line with previous reports on the sign effect on a discount rate [12]) but more inconsistent (see Table 2 for group data).

## 4. Discussion
### 4.1 Impulsivity and inconsistency in depressive patients' intertemporal choice
As far as we know, this study is the first to utilize the q-exponential discount model inspired by Tsallis' non-extensive thermostatistics, in order to examine impairments in depressive patients' decision-making. We observed that depressive patients were more impulsive in intertemporal choice *actions* and more inconsistent in temporal discounting, in comparison to healthy subjects. Our results imply that (i) depressive patients may experience preference reversal more frequently and dramatically, than healthy subjects, (ii) depressive patients may be more sensitive to potentially harmful events occurring in the extremely distant future, than healthy subjects, and (iii) although depressive patients' plans about the distant future may be forward-looking, their intertemporal choice *actions* (occurring at delay $D=0$) may be more myopic than healthy controls. Future studies should examine whether subjects who tend to make infeasible future plans are more susceptible to depression. Another possibility is that neuronal changes associated with depression (e.g., hypoactivation of serotonergic systems) may induce time-inconsistency in temporal discounting. Concerning this possibility, several studies have indicated that time-inconsistency in temporal discounting may result from nonlinear distortion of psychological time [27-29].

With respect to biophysical mechanisms underlying serotonergic modulation of time-perception in intertemporal choice, it should be noted that (i) biophysical simulation studies indicate that nonlinear psychophysical effects on sensation is mediated by electrical coupling between neurons [30] and (ii) serotonin modulates neuronal coupling [31]. Therefore, future neuroeconomic and biophysical studies should examine whether a decrease in serotoninergic activities induces both distorted time-perception and time-inconsistent discounting behavior. As noted in the introduction, findings on the relationship between serotonin and discounting have been mixed. Mobini and colleagues' study [18] demonstrated a reduction in serotonergic activities resulted in a significant increase in a discount rate; while Crean and colleagues' study [19] reported no significant effect. It should be noted that in Mobini and colleagues' study, the time-range of the intertemporal choice task was short; while Crean and colleagues' study employed longer delays, i.e., about 1 year (note that both Mobin et al.'s and Crean et al's studies did not employ the q-exponential discount function but a simple hyperbolic function [10-12], and therefore the estimated simple hyperbolic discount rates in Crean et al's study were under the influences of both relatively longer and shorter delays). Our present results of the relationship between depression (associated with a reduction in serotonergic activities) and temporal discounting may resolve the discrepancy: a decrease in serotonergic activities may increase a discount rate at short delays; while decrease a discount rate at longer delays (see Figure 2). This analysis is impossible without utilizing the q-exponential discount function.

A recent study reported that intertemporal choice for other is more inconsistent than for self [9]. It is important to examine whether intertemporal choice for other by healthy subjects involves similar neuropsychological processes to intertemporal choice for self by depressive patients. Neurochemically, a previous study implied that reduced (nor)adrenergic activities (assessed with salivary alpha-amylase levels) are associated with impulsivity in intertemporal choice [32]. It should therefore be examined whether (nor)adrenergic activities are likewise related to consistency in temporal discounting.

**4.2 Sign effects on intertemporal choice**
We observed that the signs of the outcomes (i.e., gain or loss) markedly affect both impulsivity and inconsistency in intertemporal choice (i.e., $k_q$ and $q$). Although several neuroeconomic studies examined the neural correlates of discounting delayed monetary gains [33], little is known regarding the neural processing underlying discounting of monetary loss. Future studies should examine which neuro-biophysical processes mediate the observed sign effects.

**4.3 Limitations and future directions**
As noted, the depressive patients in the present study were medicated with several types of antidepressants/mood stabilizers. Therefore, it is not completely evident that the treatments did not dramatically affect the intertemporal choice behavior. Nevertheless, our present results may not totally be attributable to the effects of antidepressants, because (i) we did not observe the effects of the types of antidepressants on the parameters in the q-exponential discount function, and (ii) BDI-II scores were significantly higher in the depressive patients than the controls, indicating that the patients were still depressed enough to elicit depression-induced alteration in intertemporal choice (we carefully scheduled our experiment so that the patients may participate during their depressive phase). In order to further resolve these issues, future studies should examine the effects of antidepressants on intertemporal choice by healthy human subjects (or animals).

**Figure legends**

**Fig. 1** Indifference points and fitted q-exponential discount function for gain (A) and loss (B). Black and red dots are indifference points at delays for healthy controls and depressive patients, respectively. Black and red curves are best-fit q-exponential discount functions for healthy controls and depressive patients, respectively. Note that in (B), the vertical axis is the unsigned (absolute) subjective value of delayed monetary loss.

**Fig. 2** Estimated q-exponential discount rate: $k_q/(1+k_q(1-q)D)$ for gain (A) and loss (B). Black and red curves are q-exponential discount rates for healthy controls and depressive patients, respectively.

Table 1. Means and standard deviations for demographic variables for depressed patients and healthy people

|  | Depressive patients | | Healthy controls | |
| --- | --- | --- | --- | --- |
|  | Mean | SD | Mean | SD |
| Sex (% men) | 55.18 |  | 40 |  |
| Age (years) | 43.83 | 2.25 | 47.6 | 3.53 |
| BDI-II | 20.52* | 2.42 | 8.93 | 1.12 |

*:Significantly larger than healthy controls ($p<0.05$). BDI-II: Beck Depression Inventory (high BDI-II scores indicate severe depression).

Table 2-1. Group data of estimated parameters in q-exponential discounting for gain

| | Depressive patients | | Healthy controls |
|---|---|---|---|
| $k_q$ (impulsivity) | 0.0006099 | > | 0.0004064 |
| $q$ (consistency) | -2.4252055 | < | -0.2132638 |

Note that larger $k_q$ and $q$ indicate more impulsive intertemporal choice at delay D=0 and more consistent intertemporal choice, respectively.

Table 2-2. Group data of estimated parameters in q-exponential discounting for loss

| | Depressive patients | | Healthy controls |
|---|---|---|---|
| $k_q$ (impulsivity) | 0.0005576 | > | 0.0002696 |
| $q$ (consistency) | -72.5 | < | -1.7333967 |

Note that larger $k_q$ and $q$ indicate more impulsive intertemporal choice at delay D=0 and more consistent intertemporal choice, respectively.

Fig.1A

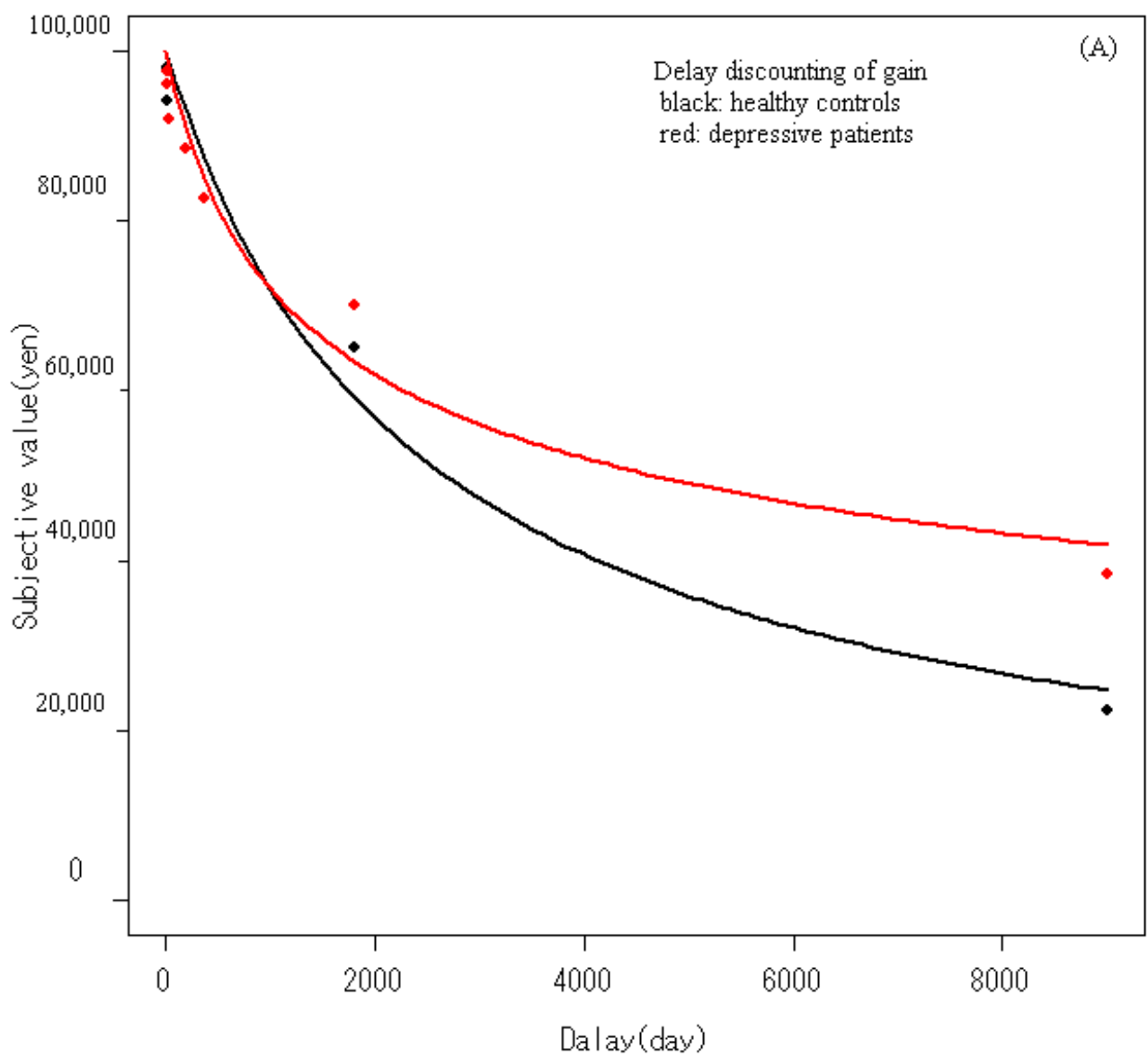

Fig.1B

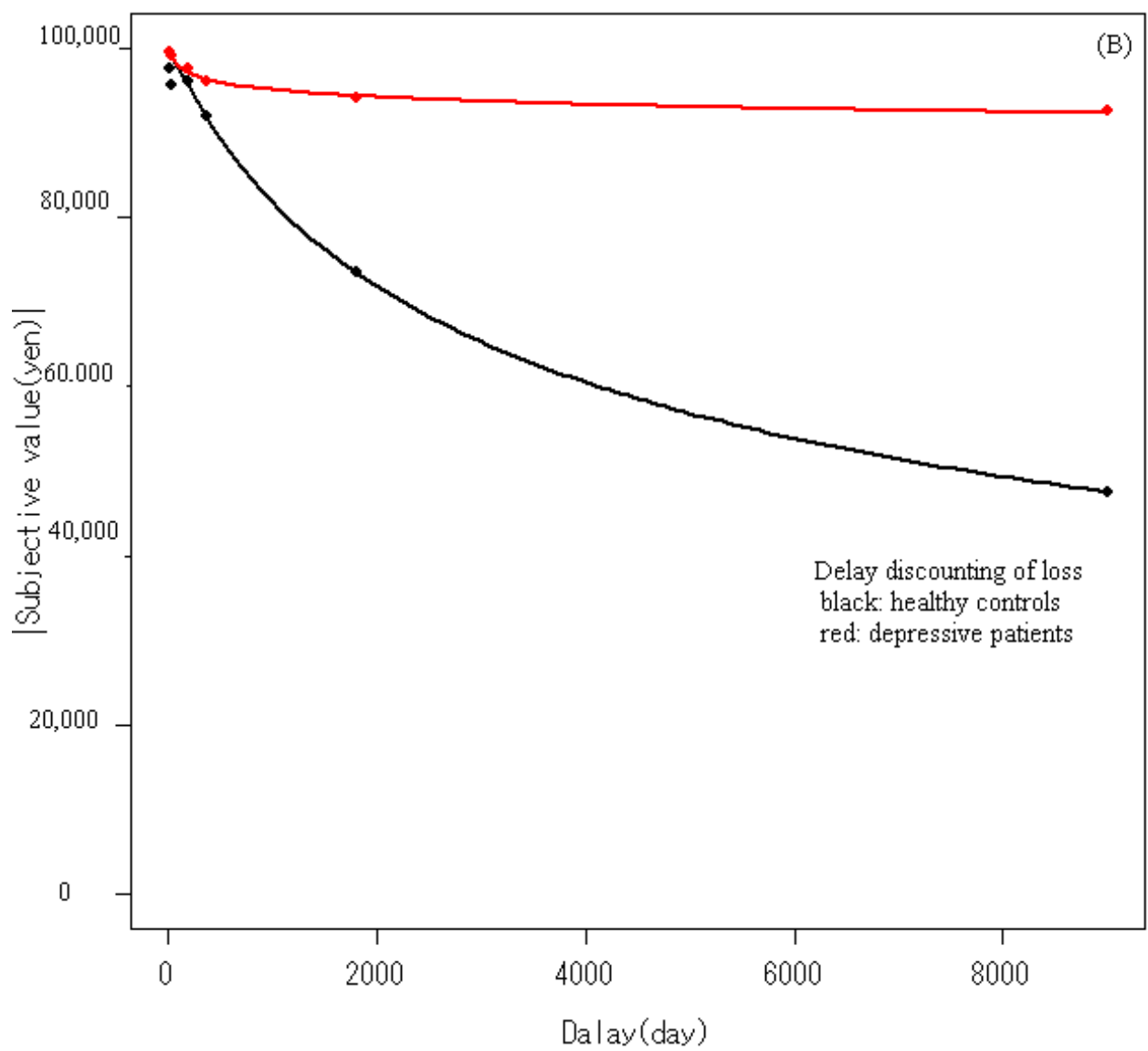

Fig2.A

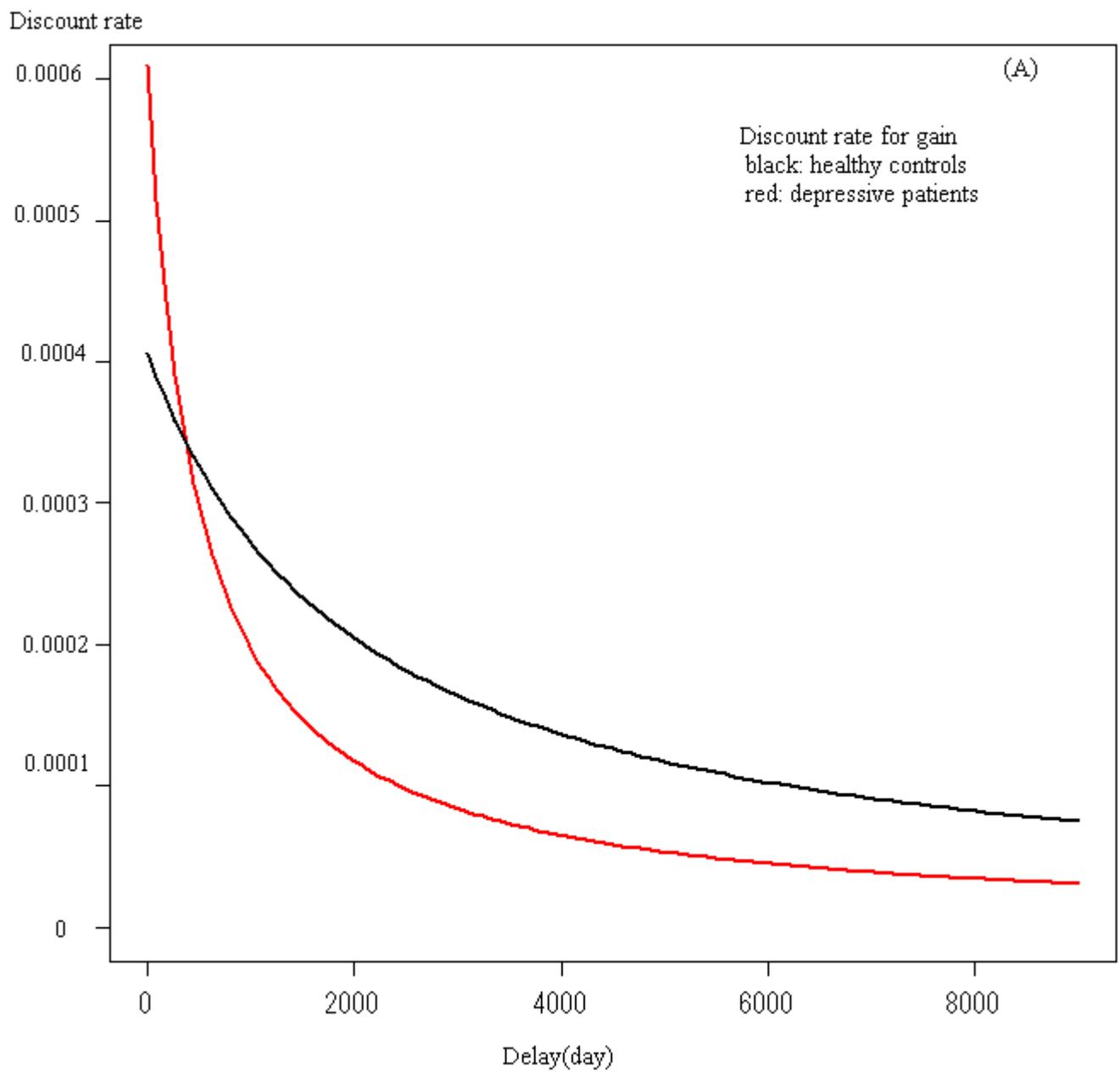

Fig.2B

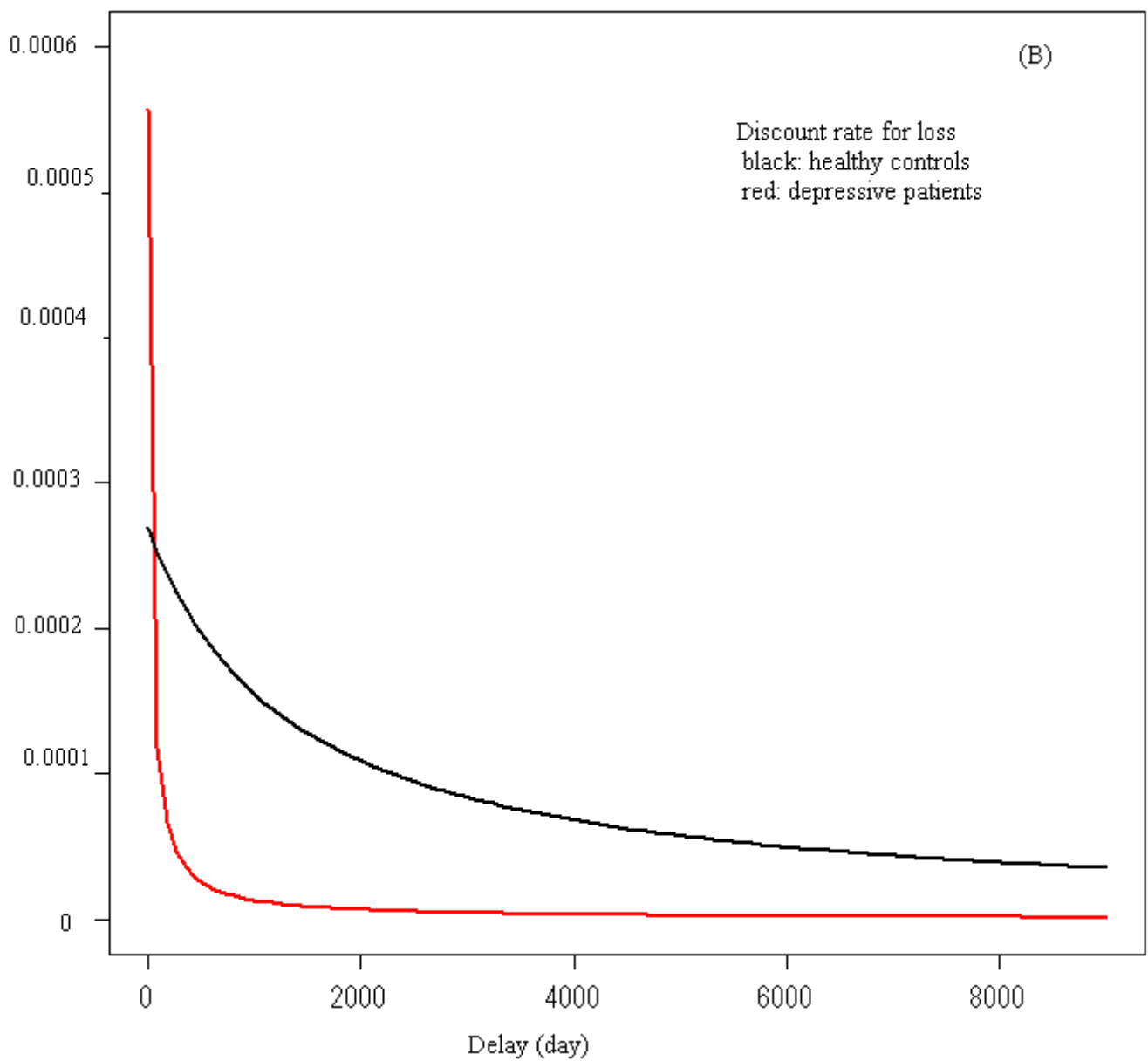